\documentclass{PoS}
\usepackage{multirow}
\newcommand{\sla}[1]%
        {{\raise.15ex\hbox{$/$}\kern-.57em #1}}

\def\al{\alpha}

\def\ga{\gamma}

\def\ka{\kappa}
\def\la{\lambda}

\def\ps{\psi}

\def\De{\Delta}

\def\lrDnu{\stackrel{\leftrightarrow}{D^\nu}}

\title{Have we tested Lorentz invariance enough?}
\ShortTitle{Have we tested enough?}

\author{\speaker{David Mattingly}\\
University of New Hampshire

        E-mail: \email{dyo7@unh.edu}}

\abstract{Motivated by ideas from quantum gravity, Lorentz
invariance has undergone many stringent tests over the past decade
and passed every one.  Since there is no conclusive reason from
quantum gravity that the symmetry \textit{must} be violated at some
point we should ask the questions: a) are the existing tests
sufficient that the symmetry is already likely exact at the Planck
scale? b) Are further tests simply blind searches for new physics
without reasonable expectation of a positive signal? Here we argue
that the existing tests are not quite sufficient and describe some
theoretically interesting areas of existing parameterizations for
Lorentz violation in the infrared that are not yet ruled out but are
accessible (or almost accessible) by current experiments. We
illustrate this point using a vector field model for Lorentz
violation containing operators up to mass dimension six and
analyzing how terrestrial experiments, neutrino observatories, and
Auger results on ultra-high energy cosmic rays limit this model.}

\FullConference{From Quantum to Emergent Gravity: Theory and Phenomenology\\
        June 11-15 2007\\
        Trieste, Italy}

\begin{document}

\section{Introduction}
Over the last century one of the most important problems in
theoretical physics has been quantum gravity and, despite efforts by
many luminary physicists, no complete model yet exists. A primary
reason why quantum gravity has remained elusive is the lack of
observational input.  Relevant observations are difficult to obtain
because of the enormous difference between terrestrial energies and
the natural quantum gravity scale - the Planck energy. The Planck
energy of $E_{Pl}=1.22 \cdot 10^{19}$ GeV is 16 orders of magnitude
larger than current accelerator energies, precluding direct probes
of Planck scale physics. However, recently it has been recognized
that some knowledge about quantum gravity can be indirectly gleaned
from low energy processes. Various models/ideas about quantum
gravity suggest that the fundamental symmetry of special relativity,
Lorentz invariance, may not be an exact symmetry but instead
violated at the Planck scale.  At low energies physics would then
presumably show tiny deviations from Lorentz invariance as well. The
exact size of these deviations is dependent on the underlying
quantum gravity model.  Apart from any quantum gravity motivations,
tests of Lorentz invariance have historically been important because
of the fundamental role Lorentz invariance plays in quantum field
theory and general relativity.

Incredibly precise and sensitive tests of Lorentz symmetry have been
performed by numerous researchers over the past two decades.  An
issue arises when considering the relevance of these tests for
models of quantum gravity, however.  While quantum gravity models
might suggest that Lorentz symmetry is not exact, there is no firm
calculated prediction from any model for the \textit{size} of the
violation. The predominant approach up to this point has been to
simply search for/constrain Lorentz violation in some low energy
framework that we hope is the right infrared limit for quantum
gravity.  The difficulty is that all of the infrared frameworks
mathematically allow for infinitesimally small Lorentz violation and
hence we can never rule out any framework a priori.  We therefore
have the unpleasant combination of incredibly precise tests of
frameworks that can never be falsified, which begs the question: are
there any reasonable spots where we might still see a signal of a
violation of Lorentz invariance, or at this point are experimental
searches simply expanding the range of validity where we know
Lorentz invariant physics works without any real expectation that we
might see a symmetry violation?\footnote{It is not the intent of
this work to argue that tests of Lorentz symmetry are ever
unimportant.  Even if the regions of theoretical interest discussed
here (and perhaps others) are eventually excluded, tests of Lorentz
symmetry are still interesting as we should always strive to extend
the limits of our physical theories. However, it is much more
important to explore areas where there are credible theoretical
ideas to be tested.}

The answer is that there are still reasonable spots. While the
infrared frameworks mathematically allow for infinitesimal amounts
of Lorentz violation, there are regions of each framework that are
preferred on physical grounds. Even better, experimentally we are
able to investigate some of these regions of theoretical interest.
Our goal in this paper is to discuss these specific experimental
possibilities.

\section{Lorentz violation in field theory}
\subsection{Lorentz violation by itself}
The most common systematic approach to studying Lorentz symmetry
violation (LV) is to construct a Lagrangian that contains the
Lorentz violating operators of interest.  The complete set of
renormalizable operators that can be added to the standard model,
the standard model extension (SME)~\cite{Colladay:1998fq}, contains
dozens of various operators made up of standard model fields and
derivative operators coupled to tensor fields with non-zero vacuum
expectation values. It is the presence of these non-zero tensors
that beaks Lorentz invariance.\footnote{We concentrate here on
coupling of Lorentz violating tensors to matter fields.  The
gravitational sector of Lorentz violating theories is much less
constrained and can generate interesting and useful phenomenology.
For a discussion see the talks by Robert Bluhm and Ted Jacobson in
this volume.}

Rather than deal with the entire SME, we can work with a simpler
model that yields the same essential physics: matter and gauge
fields couple not only to the metric, but also to a preferred frame
(c.f.~\cite{Kostelecky:1988zi,Jacobson:2000xp,Gasperini:1987nq}). We
can specify this frame by a unit timelike vector field $u^\alpha$,
the integral curves of which define the world lines of observers at
rest in the frame. $u^\alpha$ must be a field with its own kinetic
terms that induce couplings between the metric and
$u^\alpha$~\cite{Jacobson:2000xp}. Here we will only concern
ourselves with the possible couplings between $u^\alpha$ and matter
fields, specifically fermions and photons. The stable, non-trivial,
and renormalizable operators that couple $u^\alpha$ to fermions and
photons for this construction are
\begin{equation} \label{eq:fermionrotinv}
\mathcal{L}_f=\overline {\psi} (i\sla{D} - m) \psi - E_{Pl} b
u_{\mu} \overline{\ps} \ga_5 \ga^{\mu}\ps + \frac {1} {2} i c u_\mu
u_\nu \overline{\ps} \ga^{\mu} \lrDnu \ps  + \frac {1} {2} i d u_\mu
u_\nu \overline{\ps} \ga_5 \ga^\mu \lrDnu \ps
\end{equation}
and
\begin{equation}\label{eq:LVQEDphotonrotinv}
\mathcal{L}_\gamma=-\frac {1} {4} F^{\mu \nu} F_{\mu \nu} -\frac 1 4
(k_F){u_\ka \eta_{\la\mu} u_\nu} F^{\ka\la}F^{\mu\nu}
\end{equation}
respectively~\cite{Colladay:1998fq}.  $b,c,d$ and $k_F$ are the
coefficients that determine the size of any LV and each fermion
species can in principle have different coefficients.   We have
``de-dimensionalized'' the dimension three $b$ operator by the
Planck energy so that all coefficients are dimensionless. We neglect
LV terms for other gauge bosons as they will not be relevant for the
observations we will discuss, essentially because they a) have high
masses and b) do not propagate freely over long distances.

A natural question is, why do we choose the dimensionful coefficient
to be the Planck energy and not, say, the particle mass?  On one
hand, this is just a matter of convention.  However, since we are
looking for LV sourced by quantum gravity, we would expect the
quantum gravity scale to be the energy scale that controls the size
of various operators. We will see later that this assumption is
natural but leads us into serious difficulties for an experimentally
viable model for LV.

There are other terms, of course, but they are of higher mass
dimension.  If we stick with our prescription of determining the
size of the operator by $E_{Pl}$ all the higher dimensional
operators are small and irrelevant. We still list them, however, as
part of our task is to describe how these operators can become
relevant again. The complete dimension five operators have been
catalogued in~\cite{Bolokhov:2007yc}, while the dimension six
operators are not yet completely known.  Of interest to us are only
those operators that modify free particle behavior, i.e. kinetic
terms. Interaction terms of higher dimension are suppressed by the
Planck mass and lead only to very small modifications to particle
reaction rates.~\footnote{It might still be possible to see these
terms if they introduced new particle decays that do not exist in
the standard model.  However, the number of events would be
extremely small and we know of no current experiment that looks for
LV in this manner.}  The known fermion operators are

\begin{eqnarray} \label{eq:actionfermion}
\frac {1} {E_{Pl}} \bar{\psi} (\eta_L P_L + \eta_R P_R) \sla{u} (u
\cdot D)^2 \psi + \overline {\psi}\bigg{[} - \frac {1}
{E_{Pl}} (u \cdot D)^2 (\alpha^{(5)}_L P_L  + \alpha^{(5)}_R P_R) \\
\nonumber - \frac {i} {E_{Pl}^2} (u \cdot D)^3 (u \cdot \gamma)
(\alpha^{(6)}_L P_L + \alpha^{(6)}_R P_R)  - \frac {i} {E_{Pl}^2} (u
\cdot D) \Box (u \cdot \gamma) (\tilde{\alpha}^{(6)}_L P_L +
\tilde{\alpha}^{(6)}_R P_R)\bigg{]} \psi
\end{eqnarray}
where $P_R$ and $P_L$ are the usual right and left projection
operators, $P_{R,L}=(1 \pm \gamma^5)/2$, and $D$ is again the gauge
covariant derivative.  Again, we have used the Planck energy to make
all remaining coefficients dimensionless.  The currently known
photon operators are
\begin{eqnarray} \nonumber \label{eq:actionphoton}
\mathcal{L}_b=  \frac{\xi} {E_{Pl}} u^\mu F_{\mu \nu} (u \cdot
\partial) u_\alpha \tilde{F}^{\alpha \nu} - \frac {1} {2 E_{Pl}^2} \beta^{(6)}_\gamma F^{\mu \nu} u_\mu
u^\sigma (u \cdot \partial)^2 F_{\sigma \nu}.
\end{eqnarray}

It is useful to classify the set of fourteen operators above by both
mass dimension and behavior under CPT. The classification of
operators is shown below in Table~\ref{tbl:classification}.  An
$\times$ means that no operator exists with the specified properties
while a ? implies that the operators are unknown.
\begin{table}[htb]\label{tbl:classification}
\caption{Stable, nontrivial kinetic fermion and photon LV operators}
\begin{tabular}{|c|c|c|}
  \hline
 Dim & CPT Odd & CPT Even \\
  \hline
 \multicolumn{3}{|c|}{Fermions} \\
\hline

3 & $- E_{Pl} b u_{\mu} \overline{\ps} \ga_5 \ga^{\mu}\ps$ & $\times$ \\
\hline

\multirow{2}{*}{4} & \multirow{2}{*}{$\times$} & $\frac {1} {2} i c
u_\mu u_\nu \overline{\ps} \ga^{\mu} \lrDnu \ps$\\
 & & $\frac {1} {2} i d u_\mu u_\nu
\overline{\ps}
\ga_5 \ga^\mu \lrDnu \ps$\\
\hline
 5 & $\frac {1} {E_{Pl}} \bar{\psi} (\eta_L P_L + \eta_R P_R) \sla{u} (u
\cdot D)^2 \psi$ & $- \frac {1} {E_{Pl}} \overline{\psi} (u \cdot
D)^2 (\alpha^{(5)}_L P_L  + \alpha^{(5)}_R P_R) \psi$\\
 \hline
\multirow{2}{*}{6} & \multirow{2}{*}{?} & $- \frac {i} {E_{Pl}^2}
\overline{\psi}(u \cdot D)^3 (u \cdot \gamma)
(\alpha^{(6)}_L P_L + \alpha^{(6)}_R P_R) \psi$\\
& & $ - \frac {i} {E_{Pl}^2}\overline{\psi} (u \cdot D) \Box (u
\cdot \gamma) (\tilde{\alpha}^{(6)}_L P_L + \tilde{\alpha}^{(6)}_R
P_R) \psi$\\
\hline
 \multicolumn{3}{|c|}{Photon}\\
\hline
3 &  $\times$ & $\times$\\
\hline 4 & $\times$ & $-\frac 1 4
(k_F){u_\ka \eta_{\la\mu} u_\nu} F^{\ka\la}F^{\mu\nu}$\\
\hline
 5 &  $ \frac{\xi} {E_{Pl}} u^\mu F_{\mu \nu} (u \cdot
\partial) u_\alpha \tilde{F}^{\alpha \nu}$ & $\times$\\
\hline

6 & ? & $ - \frac {1} {2 E_{Pl}^2} \beta^{(6)}_\gamma F^{\mu \nu}
u_\mu u^\sigma (u \cdot \partial)^2 F_{\sigma \nu}$\\
\hline
\end{tabular}
\end{table}

For the rest of this discussion we will ignore any unknown CPT odd
dimension six operators and concentrate on the phenomenology and
constraints on the known operators.

\subsection{Constraints and the fine tuning problem}
The constraints on the operators above come from a number of
sources, from tabletop laboratory experiments to high energy
astrophysics.  Below we list the best constraints on each operator,
noting for fermions which fermion species it applies to.
\begin{table}[htb]\label{tbl:bounds}
\caption{Direct bounds on LV operators.}
\begin{tabular}{|c|c|c|}
  \hline
 Dim & CPT Odd & CPT Even \\
  \hline
 \multicolumn{3}{|c|}{Fermions} \\
\hline
3 & Neutron:$|b|<10^{-46}$~\cite{Cane:2003wp} & $\times$ \\
\hline
4 & $\times$ & Neutron: $|c,d|<10^{-27}$~\cite{Cane:2003wp}\\
\hline
5 & Electron: $|\eta_{R,L}|<10^{-5}$ ~\cite{Maccione:2007yc}& Proton: $O(10^{-1})$~\cite{Gagnon:2004xh} \\
\hline
6 & $\times$ & Proton (+extra assumptions): $O(10^{-2}, 10^{-4})$ ~\cite{Gagnon:2004xh,Alfaro:2005rs} \\
\hline
 \multicolumn{3}{|c|}{Photon}\\
\hline
3 &  $\times$ & $\times$\\
\hline
4 & $\times$ & $|k_F|<10^{-15}$~\cite{Jacobson:2002hd}\\
\hline
 5 &  $ |\xi|<10^{-7}$~\cite{Fan:2007zb} & $\times$\\
\hline
 6 & None & None\\
\hline
\end{tabular}
\end{table}

The bounds on the renormalizable and CPT odd dimension five
operators are all very tight.  To assess what these bounds mean for
theory, we first need to know what the expected size of any Lorentz
violation coming from quantum gravity might be.  As mentioned
before, there is no firm prediction from any theory of quantum
gravity that Lorentz invariance must be violated and hence no
estimate from the fundamental theory of the size of the violation.
We can, however, argue from a bottom up perspective, using the rules
of effective field theory, what the expected size should be.

The essential point is the following: If Lorentz invariance is
violated by quantum gravity, then why should our low energy world
exhibit the symmetry to \textit{any} degree, much less the
experimental situation where Lorentz invariance is at the very least
an excellent approximate symmetry?  In the physical theories we know
about, the dimensionless numbers that appear are usually of order
one. Since we have explicitly factored out the Planck scale in our
Lagrangian, which controls the scale at which quantum gravity comes
into play, the LV terms (including the renormalizable ones!) are
also therefore most naturally expected to have coefficients of order
one.

They don't of course, and therefore there must be some reason why
they are small or zero.  In light of this fact many studies turned
to the non-renormalizable operators, as these are already suppressed
by the Planck energy and hence naturally small. Concentrating solely
on these operators isn't correct though, as they will, via loop
corrections integrated up to the cutoff of $E_{Pl}$, generate the
dangerous renormalizable operators with large coefficients. The
generated coefficients are O(1) because while the non-renormalizable
operators are suppressed by $E_{Pl}$, they are involved in divergent
loop corrections.  These divergences are regulated by the cutoff of
our EFT, which is also $E_{Pl}$, and the factors of $E_{Pl}$ cancel.
This is generically known to happen in non-commutative field
theories~\cite{Anisimov:2001zc} or in field theories with a LV
regulator~\cite{Collins:2004bp}. In light of these arguments all the
CPT odd coefficients are bounded at the level of $10^{-46}$ and all
the CPT even coefficients are bounded at the level of $10^{-27}$.
Hence further improvements on existing constraints of LV are likely
irrelevant unless we can get around this extreme fine tuning
problem.

\subsection{Supersymmetry and CPT to the rescue?}
The arguments in the previous section don't rely on any particular
quantum gravity model, so they are generic as long as the quantum
gravity corrections contain LV and are describable by EFT.  There is
a hole in the argument, however - it assumes there is no new physics
between experimentally accessible energies and the Planck scale as
we integrate loop integrals with known physics. This is a large and
dangerous assumption as over the history of physics we have
encountered new physics every few orders of magnitude in energy. How
would new physics affect the above argument?  Assume for the moment,
that there exists a combination of symmetries other than Lorentz
invariance that is incompatible with all the LV renormalizable
operators and CPT odd nonrenormalizable ones.  Loop corrections
involving the higher dimension operators, instead of generating
dangerous terms, would instead cancel or be zero. In such a case,
our field theory would be more experimentally feasible, as we would
have only the CPT even higher dimensional operators to consider
which are much less tightly constrained.

If we found such a symmetry, where is it?  We don't see such an
extra symmetry at low energies, which means it should be broken at
some scale $\Lambda_b$ above 1 TeV.  Below $\Lambda_b$ this symmetry
is nonexistent, so the same EFT terms as before will exist. However,
now if we have a non-renormalizable term suppressed by $E_{Pl}$ with
an O(1) coefficient it does \textit{not} generate large
renormalizable terms.  The loop integrals will only contribute up to
the new symmetry breaking scale $\Lambda_b$, which leads to
dimension three renormalizable terms of size $\Lambda_b^2/E_{Pl}$
and dimension four terms of size $\Lambda_b^2/E_{Pl}^2$.  In terms
of our original parameterization in Table~\ref{tbl:classification},
the $b,c,d$ coefficients are now naturally of size
$\Lambda_b^2/E_{Pl}^2$!  The price to pay is the introduction of an
entirely new symmetry and a symmetry breaking mechanism.

Roughly, $\Lambda_b$ can be as low as 1 TeV and still not be seen in
direct accelerator tests on the standard model.  This would get us
to a size for $b,c,d$ of $10^{-32}$.  Current limits on $b$ are well
beyond this, however, we can set $b$ identically zero if we assume
CPT.  Lorentz invariance is usually assumed in proofs of the CPT
theorem~\cite{Greenberg:2002uu} and so if we wish to create a viable
LV model, CPT must be an assumption instead.  This assumption is, of
course, experimentally compatible with known physics.

While the above construction logically works, is there any symmetry
that can actually do the magic above?  Wonderfully enough,
supersymmetry, which has been considered for a number of other
reasons, has (almost) exactly the necessary
behavior~\cite{Bolokhov:2005cj}. The underlying reason is that SUSY
can be thought of as a field transformation symmetry, which means
that different fields can't, for example, propagate at different
limiting speeds.  The same effect occurs in two-component condensed
matter analog models for spacetime, where the limit in which the low
energy quasiparticles have the same speed is the same as the limit
where there exists a field transformation symmetry between the
quasiparticles~\cite{Weinfurtner:2006wt}. Supersymmetry forbids
renormalizable LV operators while allowing dimension five and six
operators~\cite{Bolokhov:2005cj}. Note, however, that in order to be
compatible with current limits, naively the SUSY breaking scale must
be (roughly) below 1 PeV.  This leads to an interesting method to
test for the presence of SUSY in a LV theory and a nice interplay
between low energy LV searches and high energy collider experiments.
If low energy searches for LV see a signal, that implies not only
that Lorentz invariance is violated, but also that there exists
another symmetry, which we assume is SUSY for the sake of argument,
that must exist at much lower energies. Furthermore, if we assume
that the energy scale of LV is $E_{Pl}$ the size of $c,d$ give a
prediction of the SUSY breaking scale. Finally, note that if we can
improve the bounds on $c,d$ by six orders of magnitude, to
$10^{-33}$ or so, there will be no room for the construction above
to work. Admittedly this may be a tall order, but if experiments
reach this sensitivity we can begin to rule out these split
scenarios where LV occurs at high energies but a low energy
custodial symmetry protects against experimental signatures.
Improvement by even three to four orders of magnitude will drop the
required SUSY scale to under 10 TeV, putting it close to searches
for SUSY at the LHC.

There is a significant downside for astrophysical searches for LV in
the supersymmetric case.  Supersymmetric LV theories do not
significantly modify the free field equations for high energy
particles~\cite{Bolokhov:2005cj}, i.e. O(1) $\alpha^{(5)}_{R,L}$ and
$\alpha^{(6)}_{R,L}$ terms in Table~\ref{tbl:classification} are not
present in known supersymmetric Lagrangians. Therefore experiments
searching for LV with high energy neutrinos or the
Griesen-Zatsupin-Kuzmin (GZK) cutoff for ultra-high energy cosmic
rays are not able to probe this scenario.

\section{Interesting regions in LV EFT}
The first area of interest is that mentioned
above, improving the constraints on the dimension four operators by
at least a few orders of magnitude.  The second hot spot is
establishing better direct constraints on the CPT even dimension
five and six operators to limit them well below O(1). This is
important to do, because as the above discussion shows, new low
energy physics can change the hierarchy of LV operators.  While SUSY
won't allow for the above dimension six operators, there may be some
other physics which has a similar effect. Since we are able to
simply constrain directly the CPT even dimension five and six
operators to be less than O(1), it makes sense to spend the
relatively minimal effort to do so even if we don't know what the
``new physics'' might be. What does not make sense is to continue
much further after that as the next order operators are well beyond
our experimental reach for the foreseeable future.

\subsection{Dimension six operators} We first need the field
equations for the CPT even dimension five and six operators.  For
fermions, the Hamiltonian corresponding to ({\ref{eq:actionfermion})
commutes with the helicity operator, hence the eigenspinors of the
modified Dirac equation will also be helicity eigenspinors.  We now
solve the free field equations for the positive frequency
eigenspinor $\psi$. Assume the eigenspinor is of the form $\psi_s
e^{-i p \cdot x}$ where $\psi_s$ is a constant four spinor and
$s=\pm1$ denotes positive and negative helicity. Then the Dirac
equation becomes the matrix equation
\begin{equation} \label{eq:fermiondirac}
\left(%
\begin{array}{cc}
  -m - \alpha^{(5)}_L \frac {E^2} {E_{Pl}} & E-sp - \alpha^{(6)}_R \frac {E^3} {E_{Pl}^2}  \\
E+sp  - \alpha^{(6)}_L \frac {E^3} {E_{Pl}^2}   & -m - \alpha^{(5)}_R \frac {E^2} {E_{Pl}} \\
\end{array}%
\right) \psi_s=0.
\end{equation}
We have dropped the $\tilde{\alpha}^{(6)}_{R,L}$ terms as the $\Box$
operator present in these terms makes the correction to the
equations of motion proportional to $m^2$ and hence tiny.  The
dispersion relation, given by the determinant of
(\ref{eq:fermiondirac}), is
\begin{eqnarray} \label{eq:dispfermion} \nonumber
E^2 - \frac {m} {E_{Pl}} (\alpha^{(5)}_L  + \alpha^{(5)}_R) E^2 -
\alpha^{(5)}_L \alpha^{(5)}_R \frac{E^4} {E_{Pl}^2}-(
\alpha^{(6)}_RE^3)
(E+sp)\\
-(\alpha^{(6)}_LE^3) (E-sp)=p^2+m^2
\end{eqnarray}
where we have dropped terms quadratic in $\alpha^{(6)}_{R,L}$ as
they are small relative to the first order corrections for those
terms.  Terms quadratic in $\alpha^{(5)}_{R,L}$ must be kept, as the
particle mass suppresses the linear term.

At $E>>m$ the helicity states are almost chiral, with mixing due to
the particle mass and the dimension five operators.  Since we will
be interested in high energy states, for notational ease we re-label
LV coefficients by helicity, i.e.
$\alpha^{(d)}_{+}=\alpha^{(d)}_{R},
\alpha^{(d)}_{-}=\alpha^{(d)}_{L}$. The resulting high energy
dispersion relation for positive and negative helicity particles can
easily be seen from (\ref{eq:dispfermion}) to involve only the
appropriate $\alpha^{(d)}_{+}$ or $\alpha^{(d)}_{-}$ terms.  For
compactness, we denote the helicity based dispersion terms by
$\alpha^{(d)}_{\pm}$. The difference between positive and negative
helicity particle dispersion may seem counterintuitive since the
original Lagrangian is CPT invariant. However, the helicity
dependent terms are odd in $E$ and $p$ and therefore break both
parity and time reversal, leaving the combination CPT invariant.
Note also that at energies $E>>m$, we can replace $E$ by $p$ at
lowest order, which yields the approximate dispersion relation
\begin{equation} \label{eq:dispfermionhighE}
E^2 =p^2+m^2 + f^{(4)}_{\pm} p^2 +f^{(6)}_{\pm} \frac{p^4}
{E_{Pl}^2}
\end{equation}
where $f^{(4)}_{\pm}=\frac {m} {E_{Pl}} (\alpha^{(5)}_-  +
\alpha^{(5)}_+) $ and $f^{(6)}_{\pm}= 2\alpha^{(6)}_{\pm} +
\alpha^{(5)}_- \alpha^{(5)}_+$.  Note that positive coefficients
correspond to superluminal propagation, i.e. $\partial E/\partial p
> 1$, while negative coefficients give subluminal propagation.

We now turn to photon dispersion.  In Lorentz gauge, $\partial^\mu
A_\mu=0$, the free field equation of motion for $A_\mu$ in the
preferred frame with the dimension six LV operator is
\begin{eqnarray} \label{eq:eomphoton0}
(1 - \frac {\beta^{(6)}} {E_{Pl}^2} \partial_0^2) \Box A_0=0 \\
\label{eq:eomphotoni} (\Box + \frac{\beta^{(6)}} { {E_{Pl}^2}}
\partial_0^4) A_i=0
\end{eqnarray}
where $i=1,2,3$.  With (\ref{eq:eomphoton0}) and the assumption that
LV is small, we can use the residual gauge freedom of the Lorentz
gauge to set $A_0=0$ as long as $A_\mu$ is assumed to not contain
any Planckian frequencies.  For a plane wave $A_\mu=\epsilon_\mu
e^{-i k \cdot x}$, there are hence the usual two transverse physical
polarizations with dispersion
\begin{equation} \label{eq:dispphoton}
\omega^2=k^2  +  \beta^{(6)} \frac {k^4} {E_{Pl}^2}.
\end{equation}
where we have substituted the lowest order dispersion $\omega=k$.

\subsection{Constraints} There are a number of direct constraints
that can be placed on these dispersion relations.  The best
constraints to date are those in~\cite{Gagnon:2004xh}, which place
limits around $O(10^{-2})$ on the various fermion and gauge boson
parameters from the existence of ultra high energy cosmic ray
(UHECR) protons.  In short, the existence of UHECR protons implies
that $10^{10}$ GeV protons are long-lived on astrophysical scales.
If, however, protons travel faster than the low energy speed of
light, they can emit photons via the vacuum Cerenkov effect, the
rate for which is exceedingly fast (see the appendix of
~\cite{Jacobson:2005bg} for a discussion). Similarly, depending on
the LV coefficients of various fermion species, protons can be come
unstable to electron/positron pair emission, conversion to neutrons
via positron/neutrino emission, etc.  All of these processes must be
forbidden if we see UHECR protons.  With some mild assumptions on
the behavior of the LV coefficients (for example that all gauge
bosons have the same LV coefficients), two sided bounds can be
derived via these decay processes and the parton distribution
functions (PDF's) of the UHECR protons and constituent decay
products.  The PDF's are required to calculate the net LV behavior
of the composite particles in the reactions.

With a single source species, i.e. protons, this is about the best
we can do, as all we know is that the source species must be stable.
The problem is, we already know that there is likely some other
symmetry present, so therefore there will other particles (for
example superpartners) involved in the PDF's evolved up to $10^{10}$
GeV. The constraints depend on the PDF's and so will change
depending on the symmetry, although perhaps only slightly.
Furthermore, the two sided constraints are only available with some
assumptions about how the LV coefficients behave.  As a
complementary approach, we would like to be able to construct strong
limits treating each particle as fundamental without needing
assumptions about the parton makeup or structure of the LV
coefficients.

\subsubsection{Ultra high energy cosmic ray limits}
We can construct such limits if we have two source species or if we
have a more stringent constraint than just ``this one particle
species is stable''. The Pierre Auger Observatory is able to provide
us with both possibilities.  First, Auger will be able to confirm
the location of the Greisen-Zatsepin-Kuzmin (GZK)
cutoff~\cite{Greisen:1966jv,Zatsepin:1966jv} and the detailed UHECR
spectrum around it.  The GZK cutoff is an expected cutoff in the
UHECR spectrum at $5 \times 10^{19}$ eV due to pion production from
UHECR proton scattering off the cosmic microwave background, $p +
\gamma \rightarrow p + \pi^0$. Recent results from Auger have
already confirmed the existence of the cutoff~\cite{Yamamoto:2007xj}
near the expected value, which has a number of consequences for the
LV coefficients of the CPT even dimension five and six operators.

As an example of the sensitivity of GZK physics to these operators,
we can perform a basic threshold analysis with a simplified model,
similar to what was done in~\cite{Jacobson:2002hd}.   We first
assume that parity is a good approximate symmetry and that left and
right handed protons/pions have the same LV coefficients.  Without
LV, a UHECR proton can scatter of a CMB photon with energy
$\omega_0$ and produce a pion when $E_{th}>m_\pi(2 m_p + m_\pi)/4
\omega_0$, where the threshold energy $E_{th}$ is the necessary
proton energy. LV operators change the effective mass of the proton
and pion and so will change $E_{th}$. In Lorentz invariant physics,
a $5 \times 10^{19}$ eV proton is at threshold with a $1.3$ meV CMB
photon.  If LV is such that a lower energy proton is able to produce
pions off the same region of frequency space in the CMB, one would
expect to photopion production process to be enhanced and the GZK
cutoff be lowered. Similarly, if the necessary proton energy was
raised, it would it turn raise the location of the cutoff.  Hence we
can get an estimate of the size and structure of GZK constraints by
asking how the proton threshold energies for photopion production
with a 1.3 meV photon vary in LV parameter space.  Expressed in
terms of the $f^{(6)}_p, f^{(6)}_\pi$ parameters in
(~\ref{eq:dispfermionhighE}) this yields Figure 1 for how the cutoff
location deviates from the Lorentz invariant value.
\begin{figure}[htb] \label{fig:GZK}
\includegraphics[width=5in]{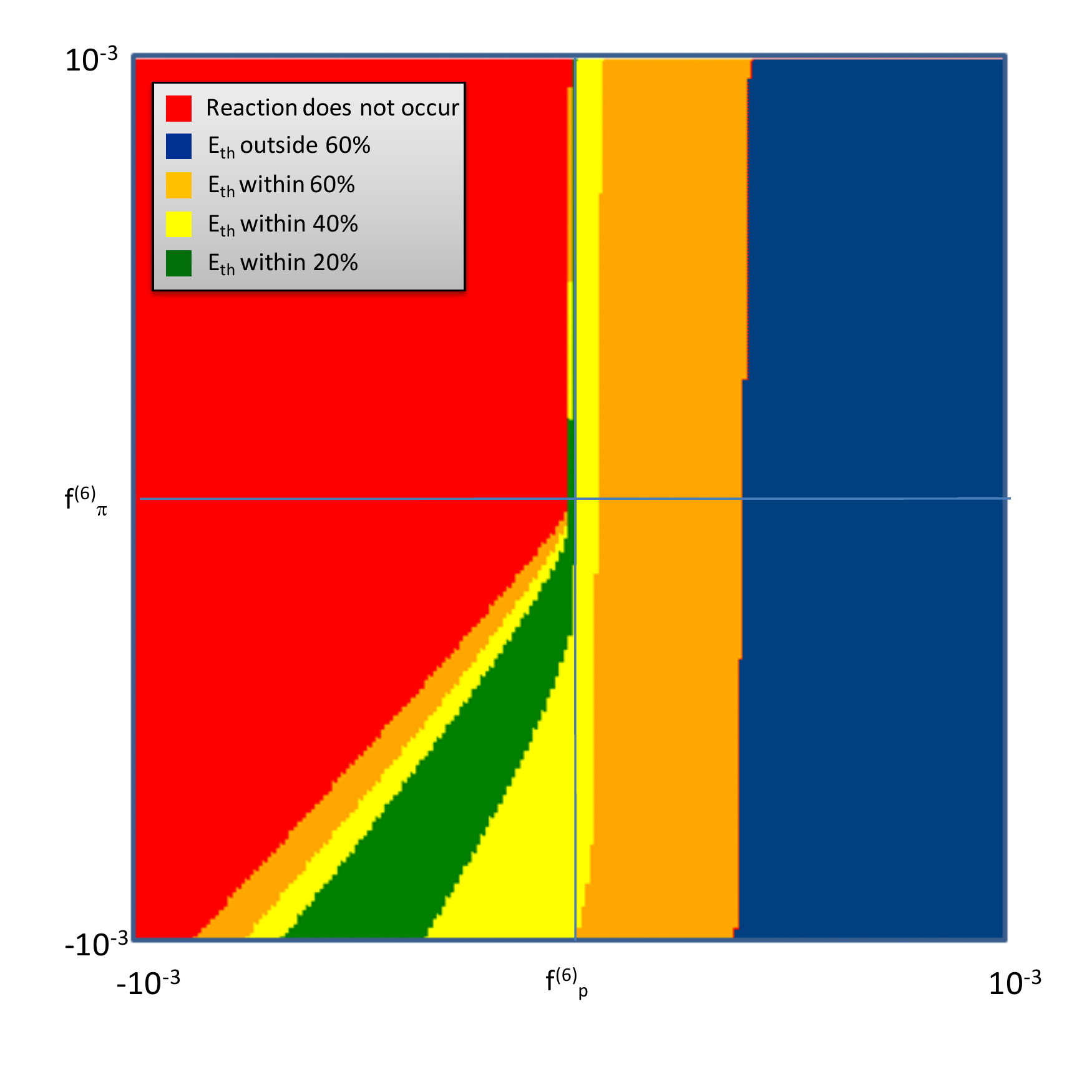}
\caption{Deviation of GZK cutoff $E_{th}$ from $5 \times 10^{19}
\rm{eV}$ as a function of Lorentz violating parameters.  To the
right of the green region $E_{th}$ is below $5 \times 10^{19}
\rm{eV}$. To the left $E_{th}$ is either greater or the reaction is
forbidden.}
\end{figure}

We see from Figure 1 that this simple requirement removes most of
parameter space.  However, we warn the reader that while the
sensitivity of the cutoff to LV parameters is evident, the situation
is more complicated than in our simple model.  The independent
coefficients for different chiralities can mask any effect, in that
opposite sign coefficients for fermions will have canceling effects.
In addition, with different fermion coefficients new effects must be
considered during propagation such as proton helicity
decay~\cite{Jacobson:2005bg} or electron-positron pair production in
the photon component~\cite{Galaverni:2007tq}. An additional, and
more important, problem is disentangling any LV source effects from
the reaction kinematics. The GZK cutoff is a deviation away from an
initial power law source spectrum. Various source models predict
different spectral indices and different composition by
species~\cite{Arisaka:2007iz}, many of which match the Auger data.
Unfortunately, there is no analysis for any source model on how LV
effects modify the spectrum. Since LV contains an energy scale near
the GZK energy at which it becomes important, it is conceivable that
the source spectrum could be a power law below GZK energies
(matching existing cosmic ray data) and change drastically above it.
It is therefore difficult to establish concrete constraints just
from the existence of the cutoff, independent of any knowledge of LV
at the source. Due to these issues, it is therefore unknown at this
time what the actual constraints on the parameter space will look
like.

We can get around many of these problems if we consider a different
way of using Auger data.  Auger can discriminate between different
cosmic ray primaries and, depending on the source dynamics, Auger
may see both proton and photon primaries~\cite{Risse:2007sd}.  This
allows us to drastically simplify our LV physics, as we no longer
need to consider source dynamics or multiple new reactions.  All we
need to require is that the photon and one chirality of proton is
stable. With LV physics, either of these particles can become
energetically unstable, in that above the threshold energy $E_{th}$
(determined again by the LV coefficients and the mass), a proton can
emit photons via vacuum Cerenkov or a photon can decay into a
proton/antiproton pair. Stable protons and photons forbids both
these reactions.  For an idea of the strength of the constraints,
let us assume again that each chirality of proton has the same LV
coefficient.  The threshold energy for these reactions to begin to
occur for O(1) $\beta^{(6)}, f_{R,L}^{(6)}$ is approximately $E_{th}
\approx (E_{Pl}^2 m_p^2)^{1/4}$, which puts $E_{th} \approx 5 \cdot
10^{18}$ eV, well below the GZK cutoff.  The timescale for either a
GZK proton to lose most of its energy or a GZK photon to decay
rapidly approaches $E_{Pl}^2/E^3$~\cite{Jacobson:2005bg} once the
energy is above $E_{th}$ which is approximately $10^{-16}$ seconds
for a $10^{19}$ eV particle. Hence the LV coefficients must be such
that neither reaction is kinematically allowed.  The parameter space
with different threshold energies is shown in Figure 2.
\begin{figure}[htb] \label{fig:protonphoton}
\includegraphics[width=5in]{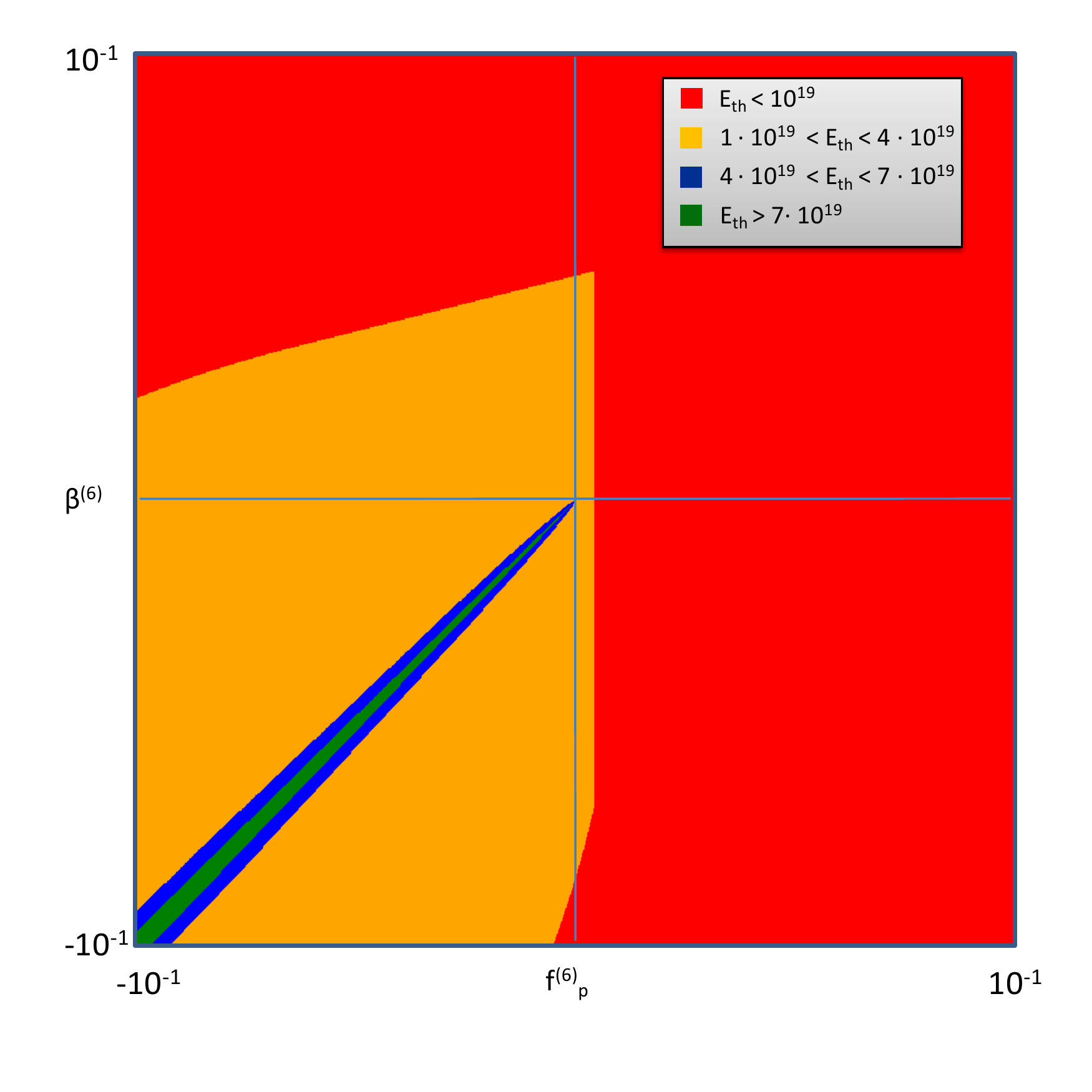}
\caption{Threshold energies $E_{th}$ for vacuum Cerenkov and photon
decay as a function of Lorentz violating parameters.  Shaded regions
are where $E_{th}$ for at least one of the reactions is in the
marked energy range.  The green area, which includes the usual
Lorentz invariant origin, contains the region where both particles
are stable.}
\end{figure}
As we can see, positive identification of different species at GZK
(or even lower) energies, can significantly restrict the allowed LV
parameter space to be very small.

\subsection{Neutrino constraints}
Due to their small mass, neutrinos provide another sensitive probe
of dimension six operators.  Remember from above that the threshold
energy for photon decay or vacuum Cerenkov involving protons was
given by $E_{th} \approx (E_{Pl}^2 m_p^2)^{1/4}$.  If we consider a
neutrino of characteristic mass $0.1$ eV the LV terms become
equivalent to the mass term at $E_{th} \approx (E_{Pl}^2
m_\nu^2)^{1/4}\approx 32 \rm{TeV}$!  This is within the observed
range of existing neutrino telescopes such as
AMANDA~\cite{Gross:2005ec}, and high statistics for this energy
range will come with next generation detectors such as
ICECUBE~\cite{GonzalezGarcia:2005xw}

Existing precision data for neutrinos is currently unable to
strongly constrain the dimension five and six CPT even operators.
For example, some of the best constraints on renormalizable
operators are provided by a combination of Super-Kamiokande
atmospheric and K2K data~\cite{GonzalezGarcia:2004wg}. To translate
these constraints into (rough) constraints on the higher dimension
operators we first need the formalism for LV neutrino
oscillations~\cite{Coleman:1998ti}. Let us consider Dirac neutrinos,
so we can use (\ref{eq:fermiondirac}). In this case, the energy
eigenstates are also the mass eigenstates. Now consider a neutrino
produced via a particle reaction in a definite flavor eigenstate $I$
with momentum $p$.  We denote the amplitude for this neutrino to be
in a particular energy eigenstate $i$ by the matrix $U_{Ii}$ where
$\sum_i U_{Ji}^\dag U_{Ii}=\delta_{IJ}$. The amplitude for the
neutrino to be observed in another flavor eigenstate $J$ at some
distance $L,T$ from the source is then
\begin{equation}
A_{IJ}=\sum_i U_{Ji}^\dag e^{-i(ET-pL)} U_{Ii}\approx \sum_i U_{Ji}
^\dag e^{-i(2E)^{-1}({m_i^2 + f^{(4)}_{-,\nu_i} p^2 +
f^{(6)}_{-,\nu_i} \frac {p^4} {E_{Pl}^2}}) L } U_{Ii}
\end{equation}
for relativistic neutrinos.  If we define an ``effective mass''
$N_i$ as
\begin{equation} \label{eq:nueffectivemass}
N_i^2=m_i^2 + f^{(4)}_{-,\nu_i} p^2 + f^{(6)}_{-,\nu_i} \frac {p^4}
{E_{Pl}^2}
\end{equation}
then the probability $P_{IJ}=|A_{IJ}|^2$ can be written as,
\begin{equation} \label{eq:oscprob}
P_{IJ}=\delta_{IJ}-\sum_{i,j>i}  4F_{IJij} sin^2 \bigg{(}\frac
{\delta N_{ij}^2 L} {4E} \bigg{)} + 2 G_{IJij} sin \bigg{(}\frac
{\delta N_{ij}^2 L} {2E} \bigg{)}
\end{equation}
where $\delta N_{ij}= N_i^2-N_j^2$ and $F_{IJij}$, $G_{IJij}$ are
functions of the $U$ matrices.  For maximal mixing between flavor
and energy eigenstates $G_{IJij}$ vanishes and the $F_{IJij}$ term
in (\ref{eq:oscprob}) reduces to
\begin{equation}
P_{IJ}=\delta_{IJ}- sin^2 \bigg{(}\frac {\delta N_{ij}^2 L} {4E}
\bigg{)}
\end{equation}

Note that above we have dropped the positive helicity terms. We are
dealing with Dirac neutrinos so there are right-handed (positive
helicity) particles. However we can ignore these as any signal will
be dominated by the left-handed neutrinos produced and interacting
via the usual standard model interactions. Therefore we are only
concerned with $f^{(4)}_{-}=\frac {m} {E_{Pl}} (\alpha^{(5)}_- +
\alpha^{(5)}_+)$ and $f^{(6)}_{-}= 2\alpha^{(6)}_{-} +
\alpha^{(5)}_- \alpha^{(5)}_+$ in (\ref{eq:nueffectivemass}).

The absolute value of the difference between $f^{(4)}_-$ terms for
muon and $\tau$ neutrinos, which acts as a change in the terminal
velocity away from $c$, is limited from the atmospheric oscillation
data collected by Super-K and K2K~\cite{Ashie:2005ik} to be less
than $10^{-24}$ in general and $10^{-27}$ for the maximal mixing
case here~\cite{GonzalezGarcia:2004wg}. This translates into the
constraint $|(\alpha^{(5)}_{\nu_\mu -} + \alpha^{(5)}_{\nu_\mu +})-
(\alpha^{(5)}_{\nu_\tau -} + \alpha^{(5)}_{\nu_\tau +}) |<10^2$ (or
larger if the neutrino mass in question is less than $0.1$ eV),
which is not particularly strong. As for the $f^{(6)}_-$ dispersion
correction, the K2K beam has an average energy of $1.3$ GeV, while
the atmospheric neutrino spectrum of Super-K has an average energy
of roughly 100 GeV~\cite{Ashie:2005ik}. The results in
~\cite{GonzalezGarcia:2004wg} use neutrinos in this energy band, so
while the direct limits on these terms have not been calculated, we
can easily overestimate the constraints on $f^{(6)}_-$ by using an
energy of 100 GeV for the neutrinos (i.e. maximizing the size of the
LV corrections). The deviation at this energy is equivalent to an
$f^{(4)}_-$ term of size $f^{(6)}_{-} E^2/E_{Pl}^2 = f^{(6)}_{-}
\cdot 10^{-34}$. Hence the constraints from Super-K and K2K on
$f^{(6)}_{-}$ are worse than $10^{7}$ and not very meaningful.

Fortunately, new detectors such as ICECUBE will dramatically raise
the neutrino energies for which there are a large observed
population of atmospheric neutrinos.  ICECUBE will see a population
of upgoing events from atmospheric neutrinos traveling through the
earth up to energies of roughly a PeV, where the earth becomes
opaque to neutrinos.  In ~\cite{GonzalezGarcia:2005xw}
Gonzalez-Garcia et. al. consider muon and $\tau$ neutrinos and
construct an observable of the number of muon events vs. zenith
angle with different values of $f^{(4)}_{\nu_\mu}
-f^{(4)}_{\nu_\tau}=2 \delta c/c$, which is the notation used
in~\cite{GonzalezGarcia:2005xw}). Atmospheric neutrinos propagating
through the earth have different path lengths as a function of
zenith angle and the variation is on the order of the diameter of
the earth, $\approx 10^7$ m.  There will be a variation in the
number of muon events with zenith angle as long as the oscillation
length between $\nu_\mu$ and $\nu_\tau$ is also of this order, i.e.
when
\begin{equation}
\frac {\delta N_{ij}^2 10^7 \rm{m}} {4E} \approx 1
\end{equation}
which we rearrange to
\begin{equation}
\delta N_{ij}^2 \approx 4 \cdot 10^{-23} \frac {E} {1 \rm {GeV}}
\rm{GeV}^2.
\end{equation}

The actual variation, taking into account attenuation and
regeneration can be found in ~\cite{GonzalezGarcia:2005xw}. For
energies around 100 TeV, the limit is $\delta N_{ij}^2 \approx
10^{-18} \rm{GeV}^2$, which puts a limit $|f^{(4)}_{\nu_\mu}
-f^{(4)}_{\nu_\tau}| < 10^{-28}$. The corresponding constraints on
the $\alpha^{(5)}_{\nu_{\mu,\tau},-}$ coefficients are therefore
still at best only of $O(10)$.  For the $f^{(6)}_-$ term, the
constraint becomes $|f^{(6)}_{\mu,-} - f^{(6)}_{\tau,-}| \approx 1$
and we finally achieve order unity constraints.  Since the $f^{(6)}$
terms scale strongly with energy, pushing the neutrino energy higher
will rapidly increase the size of the constraint.  Unfortunately, we
can only push the energy up to near a PeV, where the earth becomes
opaque to neutrinos.  At this energies, the constraints would be of
$O(10^{-2})$ . Due to the earth's opacity above a PeV it does not
appear that we will be able to go beyond this limit with a detector
such as ICECUBE.

\section{Time of flight}
\subsection{Time of flight in EFT}
Unfortunately, the cosmic ray and neutrino constraints above, while
tight, don't actually limit the absolute value of any coefficient
but instead generically limit differences of coefficients (or
functions thereof). This is obvious in the neutrino constraints and
manifests itself in Figures 1 and 2 via the qualitative form of the
constraints - open wedges in parameter space. With bounds on only
differences, Lorentz violation doesn't need to be small but only
similar for different particle species. When considering simple
changes in the terminal velocity of particles, as would be generated
by dimension four operators, this isn't an issue since the Lorentz
group doesn't specify the magnitude of the speed of light, only the
fact that some speed is invariant. Therefore constraining
differences between the terminal speeds for different species means
one is completely constraining all forms of LV at this order.  The
situation is different for the higher dimension operators.   Here,
even though the dimension five and six coefficients can be
constrained to be almost equal, LV can still be very large at high
energies. Therefore one would very much like to constrain the
absolute value of the operators, not just their differences.

It is of course possible to establish tight two sided bounds on
higher dimension operators.  CPT odd dimension five operators have
been tightly constrained on both sides from studies of the Crab
Nebula~\cite{Jacobson:2003bn,Maccione:2007yc} and a combination of
synchrotron radiation, TeV $\gamma$-ray annihilation off the IR
background, and existence of TeV photons~\cite{Jacobson:2001tu}.
Both these constraints rely on experimental confirmation that at
least three different reactions involving electrons and photons are
unaffected by LV to construct their constraints, however.  At the
very high energies needed to probe the dimension five and six CPT
even operators we no longer have this luxury and so it is difficult
to derive two-sided bounds with the threshold type reactions
considered previously.

The other method for deriving two sided bounds is to pick an
observation that only involves one LV parameter.  The simplest way
to do this is by comparing the arrival times of high energy
particles or gamma rays versus low energy gamma rays, where all
particles are emitted from the same event.  Since LV with higher
dimension operators scales with energy, the LV terms for the low
energy $\gamma$-rays are irrelevant and the arrival time delay is
effectively a function of only one parameter.  The delay $\De T$
between a low energy photon and a fermion $\psi$ with (high) energy
$E$ traveling over a time T is
\begin{equation}
\De T=  -\frac {3f^{(6)}_\psi  E^{2}} {2 E_{Pl}^{2}} T
\end{equation}
where we have neglected the $f^{(4)}_\psi$ term as it is irrelevant.

If we maximize $T$ and consider cosmogenic neutrinos at distances of
1 Gpc, we immediately see that for O(1) $f^{(6)}$ to even reach a
delay $\De T$ of one second requires energies at $10^{20}$ eV. While
such energies will likely be observed in the future by neutrino
observatories such as ANITA~\cite{Miocinovic:2005jh}, time of flight
observations are not likely to be possible.  First, one needs to
identify the sources for the flux.  For high energy neutrinos
produced as secondaries from interactions of cosmic rays with the
CMB~\cite{Stecker:1991vm}, Z-burst~\cite{Weiler:1999ny} or other
decay scenarios~\cite{Gelmini:1999ds, Birkel:1998nx}, this is
impossible. GRB's where the source can be established and the
secondary low energy signal seen have neutrino energies far too low
to constrain the LV we are considering. Hence clear two-sided bounds
on the higher dimension operators seem out of reach without imposing
additional assumptions.

\subsection{Time of flight outside of EFT}
While time of flight isn't particularly useful in constraining
higher dimension operators in EFT they are useful for proposals for
LV that do not fit within EFT and constitute our last hot spot.
There are two proposals in particular, one based on non-critical
string theory~\cite{AmelinoCamelia:1996pj}, and one coming from
``doubly special relativity'' (see~\cite{AmelinoCamelia:2003ex} for
a discussion of DSR phenomenology). Both have as one of the primary
testable phenomenological features a modified dispersion for photons
of the form $\omega^2=k^2 + \xi |w|^3/E_{Pl}$, where $\xi$ is a
coefficient universal for all photons.  There is hence a deviation
from the low energy speed of light that scales linearly with energy.
Even though this dispersion is rotationally invariant, it is
\emph{not} a dispersion law that can be constructed with a vector
field in an operator expansion. Rotationally invariant dispersion
relations where the dispersion correction is suppressed by a single
power of $E_{Pl}$ have been constructed in an EFT context, they
require the CPT violating dimension five
operator~\cite{Myers:2003fd} in (\ref{eq:actionphoton})
\begin{equation}
\frac{\xi}{E_{Pl}}u^\mu
F_{\mu\al}(u\cdot\partial)(u_\nu\tilde{F}^{\nu \al}).
\end{equation}
This operator yields a dispersion of the form $\omega^2 = k^2 \pm
\xi k^3/E_{Pl}$, where the $\pm \xi$ corresponds to right and left
circularly polarized photons. Hence there is birefringence in vacuum
for photons in addition to time of flight delays.  From the absence
of birefringence for polarized photons from afterglow of
GRB's~\cite{Fan:2007zb}, we know that $|\xi|<10^{-7}$.

Recently, the MAGIC collaboration has reported a four minute delay
in the arrival times for photons from a flare of Markarian 501 that
is compatible with a $\xi \approx -3$~\cite{Albert:2007qk}.  In an
EFT context this is meaningless, as we already have bounds $10^7$
times stronger and therefore the delay must be caused by source
effects. Conversely, if source effects are ruled out and/or a time
of flight delay of this size is confirmed for other flares, then the
conclusions will be startling. Not only would a positive result mean
that LV exists, but more drastically it would imply that standard
EFT is unable to describe at least one low energy correction from
quantum gravity! Since there are proposals for LV that don't fit
within our usual EFT framework, confirming or conclusively ruling
out this type of dispersion is an important question.   Fortunately
current experiments have the necessary sensitivity and so this
question should be answered soon.

\section{Conclusions}
LV, as illustrated by our simplified model of a LV vector field, is
very tightly constrained by current experiments.  Without custodial
symmetries LV is so tightly constrained at all orders that it seems
very likely that Lorentz invariance is an exact symmetry.  However,
a combination of CPT and supersymmetry can protect Lorentz
invariance enough that one can still have LV at the quantum gravity
scale and be compatible with experiment as long as supersymmetry is
broken at scales less than around 1 PeV.  Therefore it is still of
interest to study LV, although the models we should primarily be
interested in now must consist of both LV at the Planck and new low
energy physics.  We have illuminated here three regions that are
still of theoretical interest in studies of LV.  The first is that
of tests of the CPT even dimension four operators by low energy
experiments.  As these experiments improve, they will force the SUSY
breaking scale in a LV theory to decrease until finally it reaches
the TeV scale.  At this point SUSY will be reachable by accelerators
and we can rule out this type of split scenario.

The second region of interest is the CPT even dimension six
operators.  While supersymmetry predicts these operators to be
small, the lesson of supersymmetry is that we should not naively
trust the simple hierarchy for sizes of operators at each mass
dimension. Since we are able to directly constrain these operators
below O(1) by cosmic ray experiments such as Auger and upcoming
neutrino observatories such as ICECUBE it makes sense to do so.  The
final region of theoretical interest is in time of flight delays for
high energy particles. In EFT, time of flight observations cannot
provide us with better constraints than other experiments.  However,
there are proposals that sit outside the framework of EFT for which
time of flight delays are the only currently experimentally testable
signature.  While we personally find it implausible that the
corrections to physics at TeV energies from quantum gravity do not
fit within EFT, nature has surprised us in the past with radically
new physics.~\footnote{See, for example, quantum mechanics.}  Time
of flight observations of high energy photons from GRB's recorded by
MAGIC and other GRB observatories are able to probe these proposals
and hence this is one final region of theoretical interest.

If we test all of these regions of parameter space and find no
signal of LV, does this mean we should consider LV an exact
symmetry?  No, there are of course other forms of LV besides vector
fields and they may have different behavior. However, LV from
quantum gravity must be either more exotic (i.e. outside the realm
of effective field theory?) or carefully, and possibly unnaturally,
hidden if we do not find a signal in these regions.

\end{document}